\documentclass[pra,nofootinbib]{revtex4}

\usepackage{graphicx}
\usepackage{amssymb}
\usepackage{amsmath}
\usepackage{amsfonts}
\usepackage{epstopdf}
\usepackage{epsfig}
\usepackage{wrapfig}
\usepackage{color}

\def\ot{\otimes}
\def\>{\rangle}
\def\<{\langle}
\def\be{\begin{equation}}
\def\ee{\end{equation}}

\newtheorem{prop}{Proposition}

\begin{document}

\title{Information-thermodynamics link revisited}
\author{Robert Alicki}
\affiliation{Institute of Theoretical Physics and Astrophysics, University of
	Gda\'nsk, Poland}

\author{Micha\l{}  Horodecki}
\affiliation{Institute of Theoretical Physics and Astrophysics, National Quantum Information Centre, Faculty of Mathematics, Physics and Informatics, University of Gda{\'n}sk,  Poland}

\begin{abstract}
The so-called information-thermodynamics link created by a thought experiment of Szilard became a core of the modern orthodoxy in the field of quantum information and resources theory in quantum thermodynamics. We remind existing objections against standard interpretation of Szilard engine operation and illustrate  them by two quantum models: particle in a box with time-dependent thin potential barrier and the spin-boson model.
The consequences of the emerging superselection rules for thermodynamics and foundations of quantum mechanics are discussed.  The role of non-ergodic systems as information carriers and the thermodynamic cost of stability and accuracy of information processing is briefly discussed and compared to the generally accepted Landauer's principle.

\end{abstract} 
\maketitle
\section*{Introduction}
One of the fundamental tasks of Physics is unification of apparently different
laws describing different phenomena into a single theoretical framework. There exist many examples of successful unification: Maxwell electrodynamics, theory of electroweak interactions, statistical approach to thermodynamics, quantum theory of chemical phenomena or general relativity. In the last few decades we became a witness of a new, very ambitious, in the philosophical perspective,  unification attempt of information theory and physics. The extreme formulation due to Wheeler reads:``.. every physical quantity, every it, derives
its ultimate significance from bits, binary yes-or-no indications, a conclusion which we epitomize in the phrase, it from bit" \cite{Wheeler}.
\par
This point of view has been accepted, for instance, by the majority of physicist working in the field of quantum information and resource theory in quantum thermodynamics, sometimes giving an impression that information acquired a status of ``substance" similarly to \emph{caloric} in XVIII - XIX century physics.
Perhaps, the strongest motivation for this philosophy is provided by the Szilard approach to \emph{Maxwell's demon} concept. \\
Originally, a Maxwell’s demon \cite{Maxwell} uses manipulations at the molecular scale in order to reduce the thermodynamic entropy of a closed system, in violation of the Second Law of Thermodynamics (see \cite{Norton} for the review). More precisely, the demonic being by opening and closing a door in a  wall dividing a chamber with gas accumulates slow molecules on one side and fast molecules on the other.
\par
In the beginning of XX century the results of Einstein \cite{Einstein} and Smoluchowski \cite{Smoluchowski} finally convinced community of physicists that thermal fluctuations are observable. In principle, such fluctuations could lead to violation of the Second Law at microscopic and short-time scales. Hence, thermal fluctuations could provide a ``fuel" for a heat engine coupled to a single heat bath and employing Maxwell's demon operation principles. Indeed, a number of proposals, some of them intended for experimental implementations, has been discussed in the literature.
\par
The main message of the Smoluchowski's paper from 1912 \cite{Smoluchowski1912} was that all those  attempts to naturalize Maxwell's demon must fail.
Namely, the same thermal fluctuations which could reverse the direction of thermodynamical processes  allowing temporal violation of the Second Law act also on Maxwell's demon randomizing its operation and completely undoing the possible spontaneous decrease of entropy. Smoluchowski did not provide a general proof but studied several examples, among them the original Maxwell's proposal with the demon realized as a trapdoor controlled by a spring. Smoluchowski's arguments provide till now the most convincing exorcism of Maxwell's demon.
\par
In 1929, with the goal to clarify the idea of Maxwell's demon, Szilard proposed a model of an engine which consists of a box, containing only a single gas particle, in thermal contact with a heat bath, and a partition \cite{Szilard}.
The partition can be inserted into the box, dividing it into two equal volumes, and can slide without friction along the box. 
To extract  $k_B T\ln 2$ of work in an isothermal process of gas expansion one connects up the partition to a pulley. Szilard assumes that in order  to realize work extraction it is necessary to know  ``which side the molecule is on'' what corresponds to one bit of information.
\par
In this picture energy used to insert/remove the partition is negligible  while the subjective lack of knowledge is treated as a real thermodynamical entropy reduced by the measurement. To avoid the conflict with the Second Law of Thermodynamics it is assumed that the reduction of entropy is compensated by its increase in the environment due to dissipation of at least $k_BT\ln 2$ work invested in the feed-back protocol of work extraction. This idea initiated a never ending discussion  about the place where the external work must be invested: in measurement \cite{Brillouin:1956}, in resetting memory \cite{Landauer:1961,Bennett:2003}, or both \cite{Sagawa:2012}. Most known version is the Landauer's principle :\\
\emph{Any logically irreversible manipulation of information, such as the erasure of a bit or the merging of two computation paths, must be accompanied by a corresponding entropy increase in non-information-bearing degrees of freedom of the information-processing apparatus or its environment.}\\
In particular it means that erasure of a bit in an environment at the temperature $T$ should cost at least $k_B T \ln 2$ of work.

\par
Although, the arguments in favor of this information-thermodynamics link  seem to be generally accepted, one can find in the literature several examples of doubts and criticism.\\
\begin{enumerate}
\item{
As already noticed  by Popper and Feyerabend \cite{Feyerabend}  and more recently by one of the authors \cite{Alicki:2014},\cite{Alicki:2014a}, one can design procedures of extracting work without knowing the position of the particle.  Various designs has been proposed to eliminate the presence of observer in work extraction.}
\item{ 
Jauch and Baron \cite{Jauch} were concerned with \emph{subjectivity} of the argument based on measurement process and stressed that the \emph{potential to do work} is present even before measurement is performed.}
\item{
A well-motivated criticism of the information-thermodynamics link has been presented by Norton (see e.g.\cite{Norton}). He argued (following the reasoning of Smoluchowski) that the Szilard engine cannot yield work  without the presence of external non-equilibrium reservoir. A moving piston, massive enough to suppress thermal fluctuations, can serve as such external source of work.}
\end{enumerate}

In the present paper we would like to provide a new evidence supporting the critique of the standard paradigm. 
The paper both provides original results, as well as review some previous ones.

Our new results are about the work cost of inserting partition in Szilard engine. We discuss the case, where the partition is modeled by potential in Schrodinger equation \cite{Zurek-well}. 
Then it is known that the work cost is negligible (in particular it does not grow with $T$). Next we consider a partition modeled by physical field, and obtain that if we want to stabilize the position of the particle to be either on the left or on the right, we need to spend nonnegligible work, which must be even greater than $k_BT$. 
Moreover, we show that after insertion of the partition, the position of the particle (i.e. whether it is on the left or on the right) is correlated with two macroscopic states of the barrier. Thus, to draw work, demon is not needed anymore. Overall, in  the full cycle, the work is put and drawn in a different place than it is commonly thought. 

Based on this example, we advocate the more general view, according to which the {\it stable} information 
is not equivalent to work, i.e. that information-thermodynamics link ceases to hold for information that can be stably encoded, stored and transmitted.  Moreover we review the previous results studied in a series of mostly unpublished papers 
\cite{Alicki:2014a},\cite{Alicki:2013},\cite{Alicki:2014b}, which support the thesis that stabilizing 
information needs cost that exceeds $k_BT$, and grows along with the required stability.  We also revisit the interpretation 
of experiments that  have aimed to confirm Landauer principle, and show that the results of these indeed ingenious measurements are 
compatible with the view that we advocate in this paper. 

We further reconcile the latter view with mathematical form of Landauer principle, which is correct, near-tautological statement, by pointing out that assumptions behind the statement are not satisfied in the case of stable information - namely, the stably encoded bit, is usually not in a product state with environment (as we obtain  in our analysis of Szilard box). 

We also discuss the implication of the discrepancy between the two models of partition in Szilard box for 
foundations of quantum mechanics.

\par

\section{The physics of quantum Szilard engine}
\label{sec:engine}
Most of the criticism of information-thermodynamics link motivated by the analysis of Szilard engine is related to the energy balance in the process of inserting/removing partition. The problem is quite tricky because it involves the interaction of a microscopic system - a single molecule - with a system which must be large enough to provide a stable barrier suppressing molecule's transition from one half of a box to another one.
Moreover, partition should be also a part of environment at thermal equilibrium with the rest of the single heat bath. We discuss two types of quantum models for molecule-partition dynamics. Quantum models are simpler from the mathematical point of view and lead to some interesting questions concerning the very foundation of quantum mechanics.

\subsection{Models with potential barrier}

The process of inserting partition has been discussed in several papers in terms of a single particle in one-dimensional potential describing a fixed box with a thin potential barrier in the middle with a time-dependent magnitude.  Two examples produce the exact solutions: harmonic well \cite{Viana} or infinite square potential \cite{Joglekara} well with Dirac-delta potential at the origin parametrized as $g \delta(x)$. For the coupling $g=0$ both systems posses discrete non-degenerated energy spectrum : $E_0 < E_1 < E_2 <...$ with the corresponding even eigenfunctions  $\psi_{2k}(x)$ and odd ones  $\psi_{2k+1}(x)$, $k= 0,1,2,...$. Increasing the magnitude of the potential barrier we see that
the odd states $\psi_{2k+1}$ and their energies $E_{2k+1}$ do not change, while each even state  is deformed into ${\psi^g}_{2k}$ and its energy $E_{2k}(g)$ approaches the energy of the next odd state, i.e. $\lim_{g\to\infty}{E}_{2k}(g) = E_{2k+1}$.  The deformation is quite interesting, a sinus in the middle is formed, the state remains even but the left half of the asymptotic state $\psi^{\infty}_{2k}$ coincides with the left half of
$\psi_{2k+1}$ and its right half with the right half of $-\psi_{2k+1}$. Their linear combinations 
\begin{equation}
\phi^L_k = \frac{1}{\sqrt{2}}\bigl[\psi^{\infty}_{2k} + \psi_{2k+1}\bigr],\quad \phi^R_k = 
\frac{1}{\sqrt{2}}\bigl[\psi^{\infty}_{2k} - \psi_{2k+1}\bigr]
\label{LRstates}
\end{equation}
are completely localized in the corresponding parts of the box.
\par
According to the standard interpretation of  quantum mechanics both orthonormal basis: $\{\psi_{2k+1} ,\psi^{\infty}_{2k}; k =0,1,2,...\}$ and $\{\phi^L_k ,\phi^R_k ; k =0,1,2,...\}$ are equivalent and can be used to represent e.g. the Gibbs state in two equivalent ways. 
Moreover, the insertion of the barrier shifts the energies of even states by the factor which goes down with increasing size of the box.
Therefore, one can argue that, thermodynamically, the insertion/removal of the barrier is pretty harmless and should not essentially modify free energy of the system.
\par
On the other hand a strong enough potential barrier changes dramatically the physical properties of the system. A particle is localized always in a fixed part of the box and hence oscillates with a doubled frequency. The pressure executed by the particle on the walls is doubled.  We expect also that superpositions of left and right states, $\phi^L_k, \phi^R_{k'}$ are never observed.
This is related to the notorious problems of \emph{Schr\"odinger cat}, \emph{molecular structure}, \emph{decoherence induced superselection rules}, etc., (see e.g. \cite{Joos}). For example, as shown in \cite{Bender} the existence of such superpositions for the discussed model leads to rather non-acceptable instantaneous energy transfer on arbitrary distance. Moreover, according to the Boltzmann's definition of entropy as a logarithm of the number of \emph{accessible} states, here after insertion of partition the density of accessible states is reduced by a factor of 2 and hence a free energy of the Gibbs state is increased by $k_B T \ln 2$ providing a ``potential to work".
\par

  Thus we have a discrepancy between the model, that does not predict any substantial work cost of  inserting the barrier, and the physical intuition about such system.  
In the next Section we discuss a model of partition being a large system with many degrees of freedom, which illustrates the possible solution of the paradox of above.

\section{Spin-boson model}

The potential barrier model of partition from the previous Section is rather unphysical, in particular in the context of thermodynamical analysis. The partition must be treated as a large enough stable quantum system driven by the external forces which can execute work. It must be also at thermal equilibrium at the same temperature as the rest of environment, otherwise the argument based on the Second Law cannot be used.
\par
The proposed model consists of a particle which can occupy only two states (spin-1/2) - the left $|+\rangle$ and the right $|-\rangle$ - which are eigenstates of the discrete ``position" operator $\sigma_3$. The partition is a solid body described by a family of quantum oscillators corresponding to its different  deformation modes. The total time-dependent Hamiltonian is given by ($\hbar \equiv 1$)
\begin{equation}
H(t) =  \sum_k \omega_k a_k^{\dagger} a_k +  \lambda(t) \sigma_3 \otimes \sum_k \bigl( \bar{f}_k a_k + f_k a_k^{\dagger} \bigr) ,
\label{Ham}
\end{equation}
where $\{\omega_k\}$ are frequencies of deformation modes, $\lambda(t)$ is an overall magnitude of particle-partition coupling and $f_k$ describe the shape of partition's deformation by the presence of particle. The time-dependence of $\lambda$ describes the process of inserting or removing
of partition driven by an external agent which provides a necessary amount of work. In the Hamiltonian \eqref{Ham} we neglect all physical interaction leading to transitions between the particle states which can be described by the term 
\begin{equation}
H_1 =  \sigma_1 \otimes R ,
\label{Ham1}
\end{equation}
where $R$ is generally an environment observable. These hopping processes will be discussed later on. One should also notice that the model of partition can contain enough degrees of freedom to represent a whole (thermal) reservoir. 
\par
For a fixed value of $\lambda$ the Hamiltonian \eqref{Ham} possesses two degenerated ground states of the form 
\begin{equation}
|g\pm\rangle =  |\pm\rangle \otimes |[\pm \lambda f/\omega]\rangle,
\label{coherent1}
\end{equation}
where the global coherent state $|[\pm \lambda f/\omega]\rangle $ is a product of single-mode coherent states $|[\pm \lambda f_k/\omega_k]\rangle$ satisfying  
\begin{equation}
a_k |[\pm \lambda f_k/\omega_k]\rangle = \pm \lambda\frac{f_k}{\omega_k}|[\pm \lambda f_k/\omega_k]\rangle .
\label{coherent2}
\end{equation}
The ground state energy reads
\begin{equation}
E_g(\lambda) = - \lambda^2 \sum_k \frac{|f_k|^2}{\omega_k} .
\label{groundenergy}
\end{equation}
The important parameter is the overlap of the coherent states appearing in \eqref{coherent}
\begin{equation}
\langle[ \lambda f/\omega]|[-\lambda f/\omega]\rangle =  e^{-2\lambda^2\|f/\omega\|^2} , \quad  \|f/\omega\|^2 = \sum_k \frac{|f_k|^2}{\omega_k^2}.
\label{overlap}
\end{equation}
When partition is treated as a large system described in the thermodynamic limit the sums in \eqref{groundenergy} and \eqref{overlap} are replaced by integrals and could be infinite, typically due to the increasing contributions from the low frequency modes (infrared catastrophe).
There exists always an ultraviolet natural cut-off - Debye frequency, hence the possible  divergences are always due to ``soft phonons" . In the continuous limit one can use  ``form-factor" depending on  $\omega$ only with the power-like scaling for small $\omega$'s 
\begin{equation}
|f(\omega)| \sim \omega^{\kappa}
\label{form}
\end{equation}
For $ 0 < \kappa \leq 1$ , where $\kappa = 1$ corresponds to the Ohmic case, the ground state energy $ E_g$ is finite but the integral $\int_0^{\epsilon} |f(\omega)|^2/ \omega^2 \, d\omega$ diverges. Hence, the ground states given formally by coherent states do not overlap and therefore must be disjoint, i.e. cannot belong to a common Fock space of the bosonic system (so-called van Hove phenomenon). The most interesting regime relevant to stability of ground states is a strongly subohmic  one $  \kappa < 0$ ,
for which the ground states remain disjoint and $E_g$ diverges with infrared cut-off frequency $\omega_{min}$ 
\begin{equation}
-E_g \sim \omega_{min}^{\kappa} .
\label{energy_div}
\end{equation}
As the minimal frequency $\omega_{min}$ corresponds to the longest acoustic waves which can be supported by the system it scales like $L^{-d}$
where $L$ is a linear dimension of the system and $ d > 0 $, Therefore, the energy barrier $E_g$ separating both ground states grows with the size of the system. Such relation is always observed for systems displaying transition from quantum to classical world. For example stability of different conformations of molecules, like e.g. optical isomers, grows with their size. Moreover, ubiquitous presence of ``$1/f$- noise" phenomena in macroscopic systems suggests existence of strongly subohmic tails in spectral densities for low frequency excitations in the macroscopic world.\\ 
Notice that  exactly the same mechanisms (disjointness of different equilibrium phases in the thermodynamic limit and energy barrier between phases increasing with the system's size) explain the origin of phase transitions in quantum statistical mechanics. All that supports the claim that the localized states $\{\phi^L_k ,\phi^R_k ; k =0,1,2,...\}$ from the previous Section which belongs to physically separated ergodic components of the total system should be treated as disjoint and their superpositions as meaningless.
\par
Another approach involves dynamical origin of superselection rules for finite systems weakly interacting with large environment. A model essentially equivalent to our spin-boson one has been presented in an unpublished preprint \cite{Alicki:2013} and briefly described in one of the next Sections. This models shows the dynamical origin of metastable, macroscopically distinguishable pointer states
in the presence of heat bath. The energy barrier between such states introduces the Boltzmann factor which suppress both the error in distinguishability of those states and the tunneling rate between them. Moreover ``Schr\"odinger cat states" formed from pointer states decohere quickly to their mixtures with the rate proportional to  the square of the ``distance" between them. 
\par

\section{Thermodynamic cost of operating partition}
\label{sec:cost-partition}
When the partition is treated as a large system with many degrees of freedom one can expect that the process of operating partition involves work provided by the external driving which changes the free energy of the particle-partition system and partially is dissipated
in a form of heat. For simplicity we consider first the zero temperature case.
\par
Inserting partition for the model of the previous section is realized by increasing the coupling $\lambda(t)$ from the initial value $\lambda(0)=0$ to the final one $\lambda(t_0) = \lambda$. The initial state of the particle - partition system is assumed to be $|\pm\rangle \otimes |[0]\rangle $. Because the Hamiltonian \eqref{Ham} is quadratic the exact solution of the Schr\"odinger equation can be easily found. It is equal to  $|\Psi_{\pm}(t)\rangle =  |\pm\rangle \otimes |[\pm\chi(t)]\rangle $ where
\begin{equation}
\chi_k(t) =  \int_0^t e^{-i\omega_k (t-s)} \lambda(s) f_k \, ds = \lambda\frac{f_k}{\omega_k} - e^{-i\omega_k t} \Bigl(\int_0^{t_0} \dot{\lambda}(s) e^{i\omega_k s} ds\Bigr) \frac{f_k}{\omega_k} .
\label{evolution}
\end{equation}
Notice, that the first term on the RHS of \eqref{evolution} describes a static deformation of inserted partition while the second, time-dependent term corresponds to traveling ``acoustic wave" excited by the process of partition's insertion. Choosing for simplicity a linear slope of $\lambda (t)$ we obtain, for $t > t_0$
\begin{equation}
\chi_k(t) =  \lambda\frac{f_k}{\omega_k} - \lambda e^{-i\omega_k t} \xi_k \frac{f_k}{\omega_k} ,\quad \xi_k = \frac{1}{i\omega_k t_0}\bigl(e^{i\omega_k t_0} - 1\Bigr) , \quad |\xi_k| < 1.
\label{evolution1}
\end{equation}
The energy of the  state after insertion has form
\begin{equation}
E(t> t_0) = E_g  +  \lambda^2\sum_k |\xi_k|^2 \frac{|f_k|^2}{\omega_k} < 0 
\label{energy}
\end{equation}
Because the dynamics of the total system is a Hamiltonian one it means that $E(t> t_0)$ is equal to work performed by the external driving.
As $|\xi_k|^2 < 1$ this work is negative what can be quite easily understood. Namely, to stabilize the partition inside the box we need a kind of a ``lock" with a low enough energy of  the final ``locked state". Hence, $-E_g$ must be much larger than $k_BT$ in the case of thermal noise. During this process a part of energy  equal to $\lambda^2\sum_k |\xi_k|^2 \frac{|f_k|^2}{\omega_k} $ is carried away by ``acoustic waves" (``click of the lock") and dissipated as a heat. 

Consider now, as above, the strongly subohmic case with the energy barrier scaling \eqref{energy_div}. Then in the continuous
approximation the dissipated energy is given  by $\lambda^2 \int_{\omega_{min}}^{\omega_{max}} |\xi(\omega)|^2 \frac{|f(\omega)|^2}{\omega} $ 
and the main contribution to the integral comes from low frequencies $\omega \simeq \omega_{min}$. Therefore,  under the condition $\omega_{min} t_0 \leq 1$, $|\xi(\omega)|^2 \simeq 1$  for the relevant small frequencies and the dissipated energy is comparable to $-E_g$ and hence much higher than $k_B T$. On the other hand, the condition $\omega_{min} t_0 \leq 1$ must be satisfied because the time dependence of $\lambda(t)$ follows from the dynamics of the total macroscopic system including the partition and the control machinery. In fact  $\omega_{min}$ is the minimal frequency of the total system what implies that such system  cannot generate processes varying on the time scale longer than $\omega_{min}^{-1}$. 

Finally, to ``unlock" the partition we start from the new ground state $|\pm\rangle\otimes|[\pm \lambda f/\omega]\rangle$ and we must invest at least $-E_g >> k_B T$  of work to reach the initial state $|\pm\rangle\otimes|[0]\rangle$. Again a substantial portion of work is dissipated by ``acoustic waves".
\par 
Actually, the operation scheme of the engine proposed by Szilard is even more complicated. To transform a partition into a ``piston" a motion along the box must be ``unlocked" by the external agent with a little help of small contribution from the single-particle gas expansion. Finally, to return to the initial state several steps of ``locking" and ``unlocking" of various degrees of freedom must be used. 
One can also imagine different designs where insertion of partition cost work which is then partially recovered by its removal or the cases where both operations need work provided by external machinery.
\par
It is not difficult to generalize the model to finite temperatures. Instead of evolving pure states $|\Psi_{\pm}(t)\rangle =  |\pm\rangle \otimes |[\pm\chi(t)]\rangle $ we have two density matrices $ |\pm\rangle\langle \pm| \otimes \rho_{T}[\pm\chi(t)] $  with deformed oscillator Gibbs states at the temperature $T$, which are stable in the absence of the hopping Hamiltonian \eqref{Ham1}. The energy  balance is similar to the zero-temperature case - again an amount of work much higher than $k_BT$ is involved in the process of inserting/removing of partition. \\
The influence of hopping Hamiltonian does not change the conclusions as well. In the beginning of the insertion  process, when
 $\lambda(t) \ll \lambda$, particle is hopping between the left and right positions making unpredictable in which state is finally trapped. With increasing $\lambda(t)$  hopping probability becomes exponentially suppressed by the Boltzmann factor. At the end of removing process
hopping is switched on again what leads to complete thermalization of the system. Similarly to the pure states, the density matrices $\rho_{T}[\pm\lambda f/\omega] $  for high enough $\lambda$ must be treated as disjoint, i.e. respecting superselection rule.
\par
  
Summarizing, inserting the barrier involves two effects:
\begin{itemize}
\item[(i)] final state is a correlated state of the particle and of the macroscopic state barrier 
\item[(ii)] work much larger than $k_BT$ is dissipated during the process.
\end{itemize}
In section \ref{sec:recon} we shall discuss in more detail how these results (dis)agree with Landauer principle. 
Here let us mention, that once we are given the box with the barrier already inserted, drawing 
work does not contradict the Second Law. The barrier already measured where is the particle, and it is the barrier 
which place the role of demon. Thus one can expand the single-particle gas, conditioned on the state of the barrier, and perform $k_BT \ln 2$ 
work.

And of course, if we want to talk about complete cycle, then no positive work can be obtained (in agreement with the argument 
of \cite{Pusz} based on passivity),  as one then has to reset the barrier state. We will discuss this is more in detail in 
section \ref{sec:recon}. Here we want just to emphasize, that according to a common view, the external agent, who holds a box  with a
particle, but does not know where the particle is, cannot use the box to draw work, without gathering the knowledge. 
In contrast, here we obtain, that subjective knowledge possessed by the agent is not important, as the position of the particle has been imprinted in the state of the barrier.


However, more importantly, according to (ii) much higher than $k_B T\ln 2$ amount of work must be involved in a stable and deterministic operation of the macroscopic parts of Szilard's engine. Hence the delicate balance of quantities like $k_BT \ln2$ is completely 
overshadowed by the cost of creating system with two stable states.

  Thus, replacing the potential from sec. \ref{sec:engine} with the physical field, we obtain higher thermodynamical cost 
of processing. This suggests, that all the results, where ultimate bounds on work cost of operations were analysed
in the model where Schrodinger equation with the potential was used, should be reexamined by treating the potential physically, 
i.e. as some field.


\section{Consequences for thermodynamics and foundations of quantum mechanics}

  In previous section, we have considered an example showing how the interaction with environment can create effective 
superselection rules for quantum systems.  
For all practical purposes it is convenient to formulate a consistent description in terms of the system alone, with the influence of environment described by  mixed states obeying superselection rules and suitable effective irreversible dynamics. For a system with superselection rules the Hilbert space is decomposed into a direct sum of subspaces, $\mathcal{H} = \oplus_j \mathcal{H}_j$ corresponding to ergodic components.
\par
In our context the mathematically precise definition of ergodic component always involves some limiting procedures similarly to the definition of phases in the theory of phase transitions. Typically, a thermodynamic limit must be performed or/and cut-offs must be removed leading, for example, to infrared catastrophe. Practically, by ergodic component we mean a system with a smallest Hilbert subspace of states which are metastable with respect to a class of admissible evolutions. The states are metastable if the probability of escape per unite time from the given Hilbert subspace is negligible. The class of admissible evolutions must be defined by suitable physical requirements. For example, we begin with a fixed Hamiltonian of the system and perturb it by ``small" or ``local" perturbations.  Another method is to allow weak coupling of the system to a heat bath with restricted range of temperatures. 
\par
Remembering the above physical criteria, which must be satisfied by a non-ergodic system, we can define its state by a density matrix possessing a block diagonal structure
\begin{equation}
\rho = \oplus_j p_j \rho_j , \quad \mathrm{Tr} \rho_j = 1 ,\quad  \sum_j p_j = 1
\label{dmatrix}
\end{equation}
where the support of each density matrix $\rho_j$ is in $\mathcal{H}_j$. Here, the probabilities $p_j$ are uniquely determined and describe our \emph{subjective lack of knowledge} about system's localization. 
  As the example from previous section suggests, the capability of drawing work does not depend on this subjective knowledge, as the environment  "knows" itself which ergodic component the system is in.   Therefore, the objective, physical entropy $S_{\mathrm{ph}}$ which enters all thermodynamical expressions must be defined as an averaged entropy over all ergodic components (see \cite{Ishioka} for the very similar arguments)
\begin{equation}
S_{\mathrm{ph}}[\rho ]= k_B \sum_j p_j  S_{\mathrm{vn}}(\rho_j) = k_B S_{\mathrm{vn}}(\rho)  - k_B I[\mathbf{p}],
\label{physentropy}
\end{equation}
where $S_{\mathrm{vn}}(\sigma) = - \mathrm{Tr}(\sigma \ln\sigma)$ is the von Neumann entropy of a density matrix $\sigma$ and $I[\mathbf{p}] = -\sum_j p_j \ln p_j $ can be interpreted as \emph{information-theoretical entropy}. Indeed, any non-ergodic system which is decomposed into $n$ergodic components, can serve as a carrier of $\log_2 n$ bits of information. However,   as argued,  
the Shannon entropy $I[\mathbf{p}]$ does not contribute to the physical (thermodynamical) one: indeed, for the Szilard engine, the state of the particle after inserting partition has lower thermodynamical entropy and hence possesses a ``potential to work". 
\par

  Let us emphasize, that this does not nullify the widespraed connections between information and thermodynmics 
\cite{Goold-thermo-inf-review} including e.g. the fluctuation relations with feedback of Ref. \cite{Sagawa-Ueda}, with the important caveat, 
that the type of information which can be thermodynamically relevant is not stable. To consider most trivial example: 
the entropy of momenta of gas molecules in a container does not  represent a useful information. In contrast the position of a single molecule in Szilard box  is a meaningful information - such a box is a register that can store one bit of information. Yet, according to our view, this information is not relevant thermodynamically.  

Let us consider two particular examples. 

{\it Example 1.}  
In the paper \cite{MandalJarzynski-rotor}  the demon produces work at the expense of polluting the tape, originally filled with zeros.  
The role of demon is here played by the Markov process between the three state system and the tape (memory of the demon). 
The model is a Markov process, that is not related to physical realization,  apart from the fact, that energy change is attributed to state changes, and the detailed balance is used, to relate probability transitions with Gibbs factor. 

If we are interested in physical implementation of the model, we should ask in particular about the physics of the tape – how it is possible to have tape filled with zeros.  There are three possibilities: 
(i)	the tape is decoupled with heat bath (or the coupling is so small, that we for long time, we may consider it as effectively uncoupled);
(ii)	the tape is in equilibrium with the heat bath of low temperature;
(iii)	the bits of tape are stable, because they are ergodic components. 

The case (i) is the situation “coupling on demand” which is widely used in literature, but physically is of limited use (see also discussion in section \ref{sec:recon}). The case (ii) describes heat engine operating between two temperatures, hence it fills into standard paradigm of thermodynamics, where Landauer erasure need not be invoked to save the Second Law. Finally the situation (iii) is related to our model of Szilard engine: the zero of each bit in the tape is imprinted in the nearby environment. Our results suggest that it is then not relevant, in what states are the bits of tape on one hand, and changing bits costs more than $kT_B ln2$. 

{\it Example 2.}
In Ref. \cite{Quan-supercond} drawing work by Maxwell demon is demonstrated in terms of two superconducting qubits. One of them ($S$) interacts with hot heat bath, and the other ($D$) places a role of Maxwell demon. By means of a short circuit, the information 
from the $S$ is mapped into $D$. The information is reset in result of the contact of the system $D$ with cold bath. The situation corresponds to case (ii) described above, i.e. the qubits $S$ and $D$ are always coupled to their heat baths. So there is no stable information processed here. The information processed by means of circuit is {\it temporal} random data  from the heat bath transferred  to the the qubit $S$, and the system $D$ does not write the information in some stable memory, but it releases them to the low temperature heat bath. Thus the circuit operates on the information that is never stably stored. Also, as mentioned in Example 1, 
since we deal here with engine operating between two heat baths, and offering efficiency of Otto cycle, Landauer principle need not be invoked to save the Second Law, as simply its standard formulation is not violated to begin with. 
One can of course interprete the situation in Landauer spirit as is done in Ref. \cite{Quan-supercond},
which is an interesting way of thinking about the thermodynamics as information flow. Yet as said, it is not a flow of stable information,
but of some temporal, dynamically instable data, that immediately after emerging are sunk in the cold bath.

\subsection{  Ergodicity versus superpositions. }
The discussion on mixture of ergodic component suggests that a new postulate should be added to the standard formulation of 
quantum mechanics:\\
\emph{Superpositions of (pure) quantum  states belonging to different ergodic components do not exist, and the corresponding occupation probabilities reflect the subjective lack of knowledge, i.e. the system occupies always a given ergodic component of the Hilbert space.}
\par
Fortunately, this new postulate allows to eliminate the standard one - \emph{von Neumann projection postulate}:\\
\emph{Immediately after a measurement the state of the system is the corresponding eigenstate associated with that eigenvalue.} 
\par
Indeed, the spin-boson model used in the previous Sections can be also treated as a model of a measurement process of the observable 
$\sigma^3$. The Hamiltonian \eqref{Ham} describes the coupling of the measured spin-1/2 with the essentially macroscopic apparatus with many 
degrees of freedom represented by harmonic oscillators. The possible final states $|\pm\rangle\otimes|[\pm \lambda f/\omega]\rangle$ contain 
the stable and essentially disjoint \emph{pointer states} \cite{Zurek:2003}  $|[\pm \lambda f/\omega]\rangle$ of the apparatus strictly 
correlated with the spin eigenstates of $\sigma^3$. Then according to the new postulate the total system occupies for sure one of the states
$|\pm\rangle\otimes|[\pm \lambda f/\omega]\rangle$ and hence spin-1/2 can be certainly found in the corresponding eigenstate $|\pm\rangle$.
\par
Let us emphasize, that the proposed postulate is not fundamental, in the sense that it does not apply to quantum field theory. 
Rather it is designed to Schr\"odinger quantum mechanics which involves phenomenological potential. Infinite potential barrier 
has to be modeled on the level of quantum field theory, and the postulate emerges within  the effective picture. 
\par
 
Let us finally remark, that there is a question, whether in general stabilizing information must involve superselection rule, i.e. 
whenever we have stable bit, environment must be correlated with the values of the state of the bit. 
If self-correcting quantum memory can be built, this would be a counterexample: Indeed, the stable qubit means that superpositions between 
states are preserved, which means, that the information "which state" cannot be imprinted in environment.
At present it is an open question, whether such self-correcting quantum memory exists in three spatial 
dimensions \cite{rev-top-mem}.  In spatial dimension four, there exist stable topological quantum memories based on 
Kitaev model \cite{Dennis-top-mem,AlickiHHH-4d}, however 
these architectures are still not physical at the moment: the Hamiltonian  is ultralocal - admitting, in particular, no transport. 
 

  \section{ Proposed revision of information-thermodynamics link }  
\label{sec:correct}
  Our model for Szilard engine, analysed in section \ref{sec:engine} suggests that {\it stable} information 
is not a thermodynamic resource.  
However, there still exist interesting physical questions of the fundamental nature related to information processing. Any physical system which, by means of external controlled constraints, can be decomposed into $n$ ergodic components can serve as an universal model of information carrier with the capacity of $\log_2 n $ bits. However, the constrains are never perfect because they are always realized by finite energy barriers. In the presence of thermal noise any finite barrier is penetrable and its shape determines the life-time of encoded information at the given temperature. This life-time can be estimated using Kramers theory of reaction kinetics. For the same reasons the system localization in a given ergodic component is also never perfect and  probability tails penetrating the other components yield the error of information encoding.

\subsection{Operation cost of a switch}

The interesting problem is  the thermodynamic cost of information processing. As an elementary ingredient of information processing hardware one can take a \emph{switch} - the device which can move the  system from one ergodic component to another and back. Two models of a switch have been studied independently: a classical electronic switch by Kish \cite{Kish} and a quantum switch, described by the spin-oscillator model, in \cite{Alicki:2013},\cite{Alicki:2014b}. The latter is very similar to the spin-boson model used in this paper  and we briefly discuss its basic features.\\ 
The system consists of a spin-$1/2$ described by the standard Pauli matrices $\hat{\sigma}^k , k=1,2,3,\pm$ interacting with the harmonic oscillator with the following Hamiltonian 
\begin{equation}
{H_S} = \omega_0 {a}^{\dagger}{a} - \omega_0 g({a}^{\dagger}+{a}){\sigma}^3  ,\quad  \omega_0 , g > 0 .
\label{ham_TLSQB}
\end{equation}
The following notation for spin states, oscillator coherent states and joint spin-oscillator states is used
\begin{equation}
{\sigma}^3 |\pm\rangle = \pm |\pm\rangle, \quad {a}|\alpha\rangle = \alpha |\alpha\rangle, \quad \alpha\in\mathbf{Z},\quad |\mu;\alpha\rangle \equiv |\mu\rangle|\alpha\rangle, \quad \mu= \pm.
\label{coherent}
\end{equation}
A unitary \emph{polaron transformation} $U = e^{g( {a}- {a}^{\dagger}){\sigma}^3}$ leads to new parametrization ${a} \mapsto {b} = a-  g\sigma^3,\quad {\sigma}^k \mapsto {\tau}^{k} \equiv \sigma^k $ and transforms the Hamiltonian \eqref{ham_TLSQB} into
${H}_S=\omega_0 {b}^{\dagger}{b} - \omega_0 g^2 $ possessing degenerated ground states  $|\pm;\pm g\rangle$ with the energy $E_g = - \omega_0g^2$.\\
The spin-oscillator system (SOS) weakly interacts with the heat bath at the temperature $T$ by means of coupling linear in $\sigma^k$ and $a , a^{\dagger}$ (we disregard first environmental tunneling $\sim \sigma^1$). Using standard derivation one obtains the following quantum Markovian master equation for the density matrix of SOS 
\begin{eqnarray}
\frac{d{\rho}}{dt} = -i[{H} , {\rho}]  + \frac{1}{2}\gamma\bigl([{b}, {\rho}{b}^{\dagger}] + [{b} {\rho}, {b}^{\dagger}]\bigr)+ \frac{1}{2}\gamma e^{-\omega_0/k_B T}\bigl([{b}^{\dagger}, {\rho}{b}] + [{b}^{\dagger} {\rho}, {b}]\bigr) -\frac{1}{2}\Gamma [{\tau}^3,[{\tau}^3, {\rho}]]
\label{MME_T}
\end{eqnarray}
with the dissipation rate $\gamma $, and the pure decoherence rate $\Gamma $. Any initial SOS state $\hat{\rho}$ tends asymptotically to the mixture $p_{+}\hat\rho^{(+)} + p_{-}\hat\rho^{(-)}$ of stationary \emph{biased Gibbs states}
\begin{equation}
\rho^{(\pm)} =  \bigl(1-e^{-\omega_0/k_B T}\bigr)|\pm\rangle\langle\pm|\,e^{-\frac{\omega_0}{k_B T}{b}^{\dagger}{b}} = \bigl(1-e^{-\omega_0/k_B T}\bigr)|\pm\rangle\langle\pm|\,e^{-\frac{\omega_0}{k_B T}({a}^{\dagger}\mp g)({a}\mp g)}
\label{gibbs_biased}
\end{equation}
For $g >> 1$, the harmonic oscillator components  $\rho^{(\pm)}_{\mathcal{P}} =  \bigl(1-e^{-\omega_0/k_T}\bigr)e^{-\frac{\omega_0}{k_B T}({a}^{\dagger}\mp g)({a}\mp g)}$ of SOS state \eqref{gibbs_biased} represented the semi-classical \emph{pointer states} which encode a bit of information with the error given by the overlap defined as the transition probability (see \cite{Scutaru:1998} for the derivation) 
\begin{equation}
\epsilon = \mathrm{Tr}\Bigl(\sqrt{\sqrt{\hat\rho_{\mathcal{P}}^{(+)}}\hat\rho_{\mathcal{P}}^{(-)}\sqrt{\hat\rho_{\mathcal{P}}^{(+)}}}\Bigr) =  \exp\Bigl\{-4 D^2\tanh\bigl(\frac{\omega_0}{2k_B T}\bigr)\Bigr\} .
\label{trans_prob}
\end{equation}
This  probability of error in the process of pointer states discrimination (measurement error)  can be rewritten as a ``Boltzmann factor"
\begin{equation}
\epsilon =  \exp\Bigl\{-\frac{\bar{2|E_g|}}{k_B\Theta}\Bigr\} , \quad k_B\Theta= \frac{\omega_0}{e^{\omega_0/k_B T} -1} + \frac{\omega_0}{2}.
\label{highT}
\end{equation}
Here $2|E_g|$ is the energy gap between pointer states and their excitations by spin reversal, and  $\Theta \equiv \Theta[T,\omega_0]$ is the effective temperature which in the semiclassical regime, i.e. for $\frac{\omega_0}{k_B T} << 1$, is equal to the temperature $T$, and in the low temperature regime includes quantum fluctuations. \\
Description of tunneling process between the SOS states generated by the coupling to the heat bath through $\sigma^1$ is quite complicated and involves the derivation of the additional term to the Master equation \eqref{MME_T}, followed by a sequence of approximation. The leading order expression for the tunneling rate $\Gamma_{tun}$ contains again the Boltzmann factor and the pure decoherence rate $\Gamma_1$ for a $\sigma^1$ basis, computed for the spin alone coupled to the same bath 
\begin{equation}
\Gamma_{tun} \simeq \Gamma_1  e^{-\frac{2|E_g|}{k_B \Theta}}.
\label{tunrate_A3}
\end{equation}
Taking into account that to overcome the energy barrier $2|E_g|$ one needs a comparable amount of work one can propose the following link between information and thermodynamics:\\ 
\emph{The minimal work $W_{min}$ needed to encode or change a bit of information with an error probability $\epsilon$ under the influence of combined thermal and quantum noise at the effective noise temperature $\Theta$ is given by
\begin{equation}
W_{min} \simeq  k_B\Theta \ln{\frac{1}{\epsilon}}.
\label{minwork}
\end{equation}
}
It it interesting that the formula \eqref{minwork} has been obtained by Brillouin a long time ago \cite{Brillouin:1956}, for a particular example of measurement involving photons and with $\Theta = T$. In that case $W_{min}$ was a minimal work necessary to perform this measurement with the accuracy $\epsilon$.

\subsection{The minimal cost of long computation}
\label{sec:minwork}

The formula \eqref{minwork} is in agreement with our everyday experience, the more accurate and stable is information carrier the more work has to be used for information processing - ``engraving on a stone tablet" costs much more than making `` a mark on paper".
It allows also to answer a practical question: \emph{What is the minimal work needed to perform an algorithm which consists of $N$ elementary logical steps?}
\par
The details of estimation based on the quantum switch model can be found in \cite {Alicki:2013}. The final formula valid for $N\to\infty$ and high temperatures ($\Theta \simeq T$ ) reads 
\begin{equation}
\bar{W}_N \simeq k_B T  N  \bigr(  \ln N +  \ln \frac{1}{\delta}+  \ln \frac{1}{\kappa}\bigr) ,
\label{comp_error1}
\end{equation}
where  $\delta$ is an overall probability of failure and  $\kappa < 1$ is a ratio of decoherence time to dissipation time computed for the unprotected information carrier using the above model of a switch.  One can notice the differences between the formula \eqref{comp_error1} and the prediction based on the Landauer's principle $\bar{W}_N^{(L)} \simeq k_B T  N \ln 2$. The additional term $\ln N$ is necessary to keep the overall error probability fixed. For example, a modern supercomputer performing $10^{16}$ logical gates per second and working for a day executes an algorithm with $N \simeq 10^{21}$. Such large $N$ allows to neglect the terms dependent on $\delta$  and $\kappa$. Then, at the room temperature $300 K$ the total minimal work  $\bar{W}_N \simeq 10^2 J$ what is still much lower than the actual energy consumption, of the order of $10^{10} J$, but essentially higher than $\sim 1 J$ predicted by the Landauer's formula.

\subsection{The meaning of experimental results} 
\label{sec:exp}

One can find few papers 
\cite{Toyabe,Lutz,Jun,Hong} the authors of which claim to test experimentally Landauer's principle. However, a closer look at those,
in fact, ingenious experiments shows that at least in the first three papers the authors confirmed only that increasing or reducing a volume occupied by a Brownian particle reduces or increases  its free energy by $k_B T \ln 2$. The change of the volume is performed by controlled deformation of the external potential either in the form of double well or staircase. The potentials are generated by the external macroscopic devices which by each operation step consume energy of many orders of magnitude higher than $k_BT$. However this cost is not included in the energy balance. As explicitly, but incorrectly, stated by one of the authors \cite{Toyabe}: ``.. in the ideal case, energy to place the block can be negligible ". In reality, ``placing a block'' (i.e. modifying external potential) costs at least several $k_BT$ of work. Similarly, in the paper \cite{Lutz} only the energy dissipated by the Brownian particle is taken into account, while the work invested into control of the varying double-well potential, corresponding to the partition is disregarded (see also \cite{Jun} for more higher precision experiment).  
\par
The precise experiment of \cite{Hong} deals with a single-domain magnetic dot (nanomagnet) as an information carrier. Magnetization along the ``easy axis" is used to encode a bit of information and the ``hard axis" is related to degrees of freedom providing the potential barrier of ``partition". In fact, using the Holdstein-Primakoff theory of ferromagnets one could  obtain a description of this system similar to our spin-boson model. In the theoretical analysis of the experimental data the authors compute work performed by time-dependent external magnetic fields from the areas of two hysteresis loops.  They treat a \emph{difference} of two contributions, indeed close 
to $k_BT \ln 2$, as the cost
of the logical operation. However, any process with hysteresis is irreversible and accompanied by \emph{internal friction}. As illustrated above by the spin-boson model even if an operation step should in principle extract work from the partition, most of this work is dissipated by emission of energy quanta (magnons in the case of ferromagnet) Therefore, the work dissipated in one hysteresis loop cannot feed the other irreversible process associated with the second hysteresis loop. Actually, the total cost should be rather estimated by the \emph{sum} of areas of the loops which is essentially higher than $k_BT$.

\section{Reconciliation with Laundauer principle}
\label{sec:recon}

Having said all that we would like to make make sure that our claims are compatible  with the quantum information approach, where the Landauer principle is accepted as a simple to prove tautology, in seemingly completely general setup. Let us present a variant of 
Reeb and Wolf \cite{Reeb} formulation and proof of Landauer principle.  
For arbitrary (not necessarily equilibrium) state $\rho$ on the system with Hamiltonian $H$, we define free energy as
\be
F(\rho)= Tr (H \rho) - k_B T S_{\mathrm{vn}}(\rho).
\ee

\begin{prop}
\label{prop:Landauer}
Consider a system $S$ and a reservoir $R$, with the initial state $\rho^{in}_{RS}=\rho^\beta_R \ot \rho_S$   where $\rho_R^\beta$ is Gibbs state ($\beta = 1/k_B T$),
and the final state denoted by $\rho^{out}_{RS}=U \rho^{in}_{RS} U^\dagger$, with $U$ being an arbitrary unitary. Define heat consumed by a reservoir by change of its energy
\be
\Delta Q=Tr(H_R (\rho_R^{out} - \rho_R^{in})).
\ee
Then we have 
\be
\Delta Q \geq \Delta F_S
\ee
where 
\be
\Delta F_S= F(\rho_S^{out})-F(\rho_S^{in}).
\ee
\end{prop}
{\bf Proof.}
For the total system $RS$ the free energy is equal to average energy, so that 
$\Delta W_{RS} =  \Delta F_{RS}$. We now use formula $ F(\rho)=F(\rho_\beta)  + k_B T S(\rho|\rho_\beta)$, where $S(\rho | \sigma ) \equiv Tr (\rho \ln \rho  - \rho \ln \sigma)$,
which implies that 
\be
\beta\Delta F_{RS} = S(\rho^{out}_{RS} | \rho^\beta_{RS}) - S(\rho^{in}_{RS} | \rho^\beta_{RS}).
\ee
Since $\rho^{in}_{RS}=\rho_R^\beta \ot \rho^{in}_S$, and $\rho^\beta_{RS}=\rho_R^\beta \ot \rho^\beta_S$ by monotonicity of 
relative entropy we obtain 
\be
\beta\Delta F_{RS} \geq S(\rho_S^{out}|\rho^\beta_S) -  S(\rho_S^{in}|\rho^\beta_S) = \beta \Delta F_S.
\ee
This ends the proof. 

The above result seems to be completely general - we do not assume any model, any particular coupling. Moreover, it seems to 
put the thermodynamical entropy (related to ergodicity) and information entropy - related to lack of knowledge - on the same footing.
Indeed, the amount of work does not depend on the origin of the entropy. There is one assumption though that restricts 
the scope of applications of so formulated Landauer principle. Namely it is assumed that the initial state is product with the bath.
This assumption is used in the vastly developing domain of application resource theoretic approach to thermodynamics 
\cite{Janzing,HO}

\begin{figure}[ht]
	\centering
	\includegraphics[width=8cm]{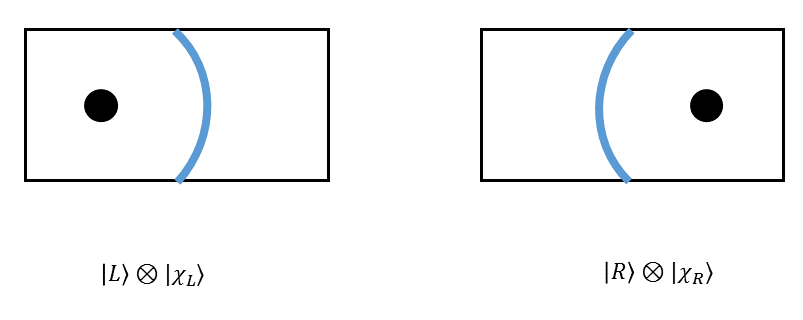}
	\caption{Partition as a Maxwell Demon: the shape of the partition depends on the location of the particle }
	\label{fig:bow}
\end{figure}

However, it clearly does not apply to   our model of the   Szilard engine for a simple reason: the states "left" and "right" are strongly, almost perfectly  correlated with states of  partition. We can imagine, that the partition bows on the left or right depending where is the particle, due to a pressure of the particle, see Fig. \ref{fig:bow}.

Thus the partition is the Maxwell demon itself, and therefore the potential of drawing work from such a system does not change 
when some additional observer will get to know where the particle is.  
This is related to emergence of objectivity studied in many models \cite{Zurek, object}.  

\begin{figure}[ht]
	\centering
	\includegraphics[width=0.7\textwidth]{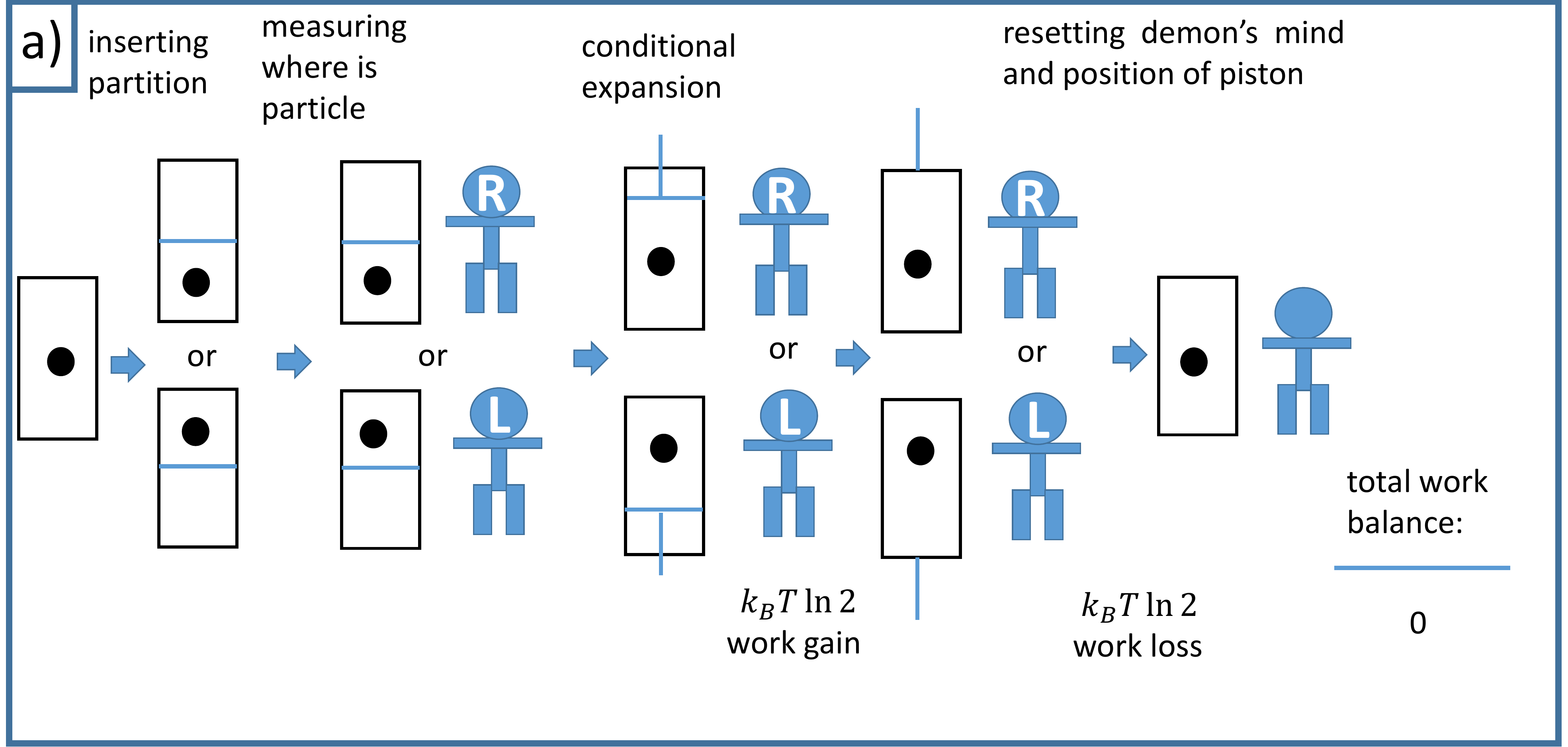}
	\includegraphics[width=0.7\textwidth]{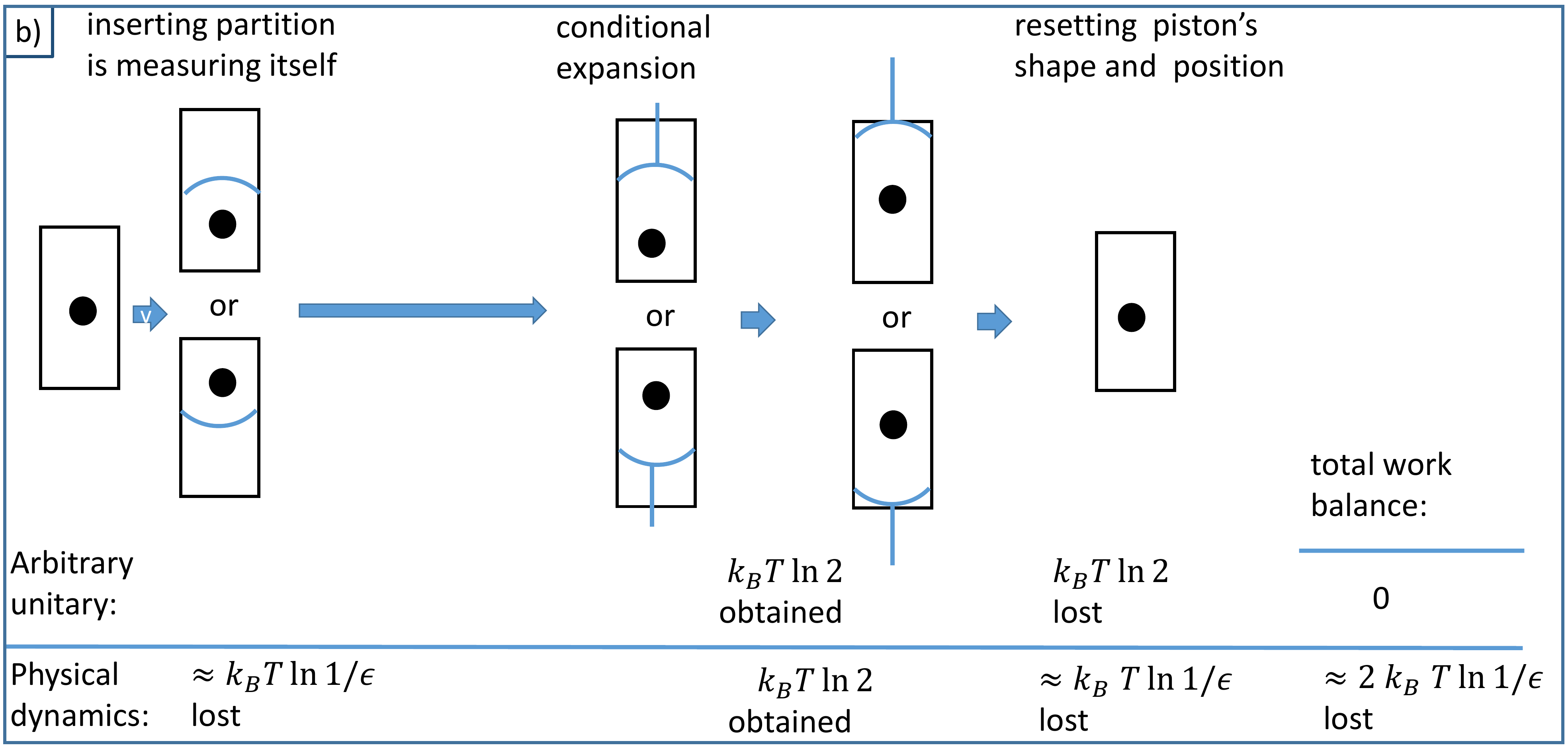}
	\caption{Comparison of standard description of the operating cycle of Szilard engine with the proposed physical description. Note, that in 
	the idealized setup -- where arbitrary unitary is allowed -- the cost of erasing shape and position of the barrier is just $k_BT\ln 2$, because the shape is correlated with the position.}
	\label{fig:szilard}
\end{figure}

If we now allow to apply an arbitrary unitary to the Szilard engine, as is allowed by the assumptions of the above theorem, we can clearly reset the value of the bit, {\it without measuring  where the particle is}. 
Thus, while the commonly accepted  application of Landauer principle to Szilard model is incorrect  -- as it says that to draw work 
the observation must be done to know  where is the particle -- still, 
in essence Landauer principle applies, only that the partition has become a demon 
(in Fig. \ref{fig:szilard} we compare those two situations).

However, application of such an operation seems unphysical. Physically, the measurement performed by the partition 
costs more than $k T_B\ln 2$, as evaluated in \eqref{minwork}. 
Also, the step of erasure, we believe, should require the amount of work  to be dissipated of the same order as 
the cost of gate operation, as to reset bit, we need to destabilize it and stabilize again, which 
will incur the cost of \eqref{minwork}, as indicated on the bottom of Fig. \ref{fig:szilard}.  

There is a question, what is the scope of application of the Landauer principle as formulated in Proposition \ref{prop:Landauer}. 
Clearly, it is applicable in a  situation, where a system can be coupled to and decoupled from environment on demand.
Indeed, then the state of the system is stable just because it is not always coupled to reservoir, and there is no additional cost 
of stabilizing it. Moreover, in such situation the total state can be product with environment: indeed, the correlations described in Szilard model arise, because we want stable state despite it being coupled to the reservoir. However the situation of coupling on demand is not likely to be found in practice, as it requires some resources to keep system uncoupled to heat bath - usually it is done by cooling that requires massive energy expenditure - incomparably big with $kT \ln 2$, as we have mentioned while discussing experiments claiming to achieve Landauer's limit.   It is not excluded though, that for concrete physical setups and in some time scales, we may treat some systems as effectively not coupled to the heat bath, which makes room for applications of the standard Landauer paradigm.  

To summarize, we believe that there is a universal cost of erasure, however, it is not given by $kT \ln 2$, but it is generically much greater, depending  on stability we want to have. Moreover, it does not come from the basic relations like unitarity of total dynamics
and monotonicity of relative entropy, but comes from the physics of interactions. Thus the real {\it devil is in details}: if we want to estimate in more generality the cost of erasure, we need to study models, which describe situation on a more fundamental level than Schrodinger equation with some a priori potential.  

\section{Concluding remarks}

  The  discussion of above suggests the following picture. If by information we mean the stable information i.e. one that can be reliably stored and transmitted, rather than information carried by fast changing random configurations of physical systems, we expect that the following holds:

I) \emph{Stable  information is an abstract entity}\\
 Stable Information $I(\mathbf{p})$, measured for simplicity in \emph{nats} ($\log_2 e \simeq1.443$ bits), is a 
\emph{disembodied abstract entity independent of its physical carrier}. In particular, multiplying $I$ by the Boltzmann constant $k_B$ or thermal energy  $k_B T$ does not yield thermodynamical entropy or meaningful energy which should be taken into account in the energy-entropy balance related to the First and Second Law of Thermodynamics. The quantities like  $k_B\times I$ or $k_B T\times I$ possess no physical meaning. In particular, stable information is not convertible into energy. Similar statement illustrated by a number of classical measurement schemes can be found in \cite{Kish:2018}.

II) \emph{Only information carriers are physical}\\
It is true that, as Landauer wrote :\emph{ "[Information] is always tied to a physical representation. It is represented
by engraving on a stone tablet, a spin, a charge, a hole in a punched card, a mark on paper, or some
other equivalent. This ties the handling of information to all the possibilities and restrictions of our real
physical word, its laws of physics and its storehouse"}.\\
However, the legitimate questions concern the physical properties of \emph{information carriers} like  "stone tablet, a spin, a charge, a hole in a punched card, a mark on paper", but not the information itself.

III) \emph{How to encode information?}\\
To encode information one needs physical systems possessing a number of distinguishable and {\it stable} with respect to thermal and quantum noise states. Therefore, for example, gas of atoms may possess a well-defined entropy but does not encode any information. The distinguishability of states can be quantified  in terms of their overlap, well-defined for classical and quantum systems. Only well-distinguishable states can be cloned with a high accuracy and  protected against noise for long times \cite{Alicki:2006}. Because our present day computers are based on macroscopic elements, which are highly stable on human time-scale and are applied to moderately large inputs and outputs appropriate again to the human scale, we can still disregard, to a large extend, physical limitations and apply abstract complexity theory to efficiency problems. \\

  We were not of course able to give a general no-go result, in favor of the above picture. However we believe that our paper will be stimulating to consider more detailed models in order to verify whether the presented view  indeed is the correct description of the
relations between thermodynamics and stable information. 

{\bf Acknowledgement.} MH is supported by National Science Centre, Poland, grant OPUS 9. 2015/17/B/ST2/01945.

\end{document}